\documentclass{ectj}

\usepackage{amsfonts,amssymb,graphics,epsfig,verbatim,bm,latexsym,amsmath,url,amsbsy}
\usepackage{tikz}
	\usetikzlibrary{arrows,positioning} 
\usepackage{subcaption}
\usepackage{bbm}
\usepackage{booktabs}
\usepackage{setspace}

\newtheorem{definition}{Definition}

\renewcommand{\thesection}{\arabic{section}}
\renewcommand{\theequation}{\arabic{section}.\arabic{equation}}

\newcommand\independent{\protect\mathpalette{\protect\independenT}{\perp}}
\def\independenT#1#2{\mathrel{\rlap{$#1#2$}\mkern2mu{#1#2}}}

\title[Double Machine Learning and Automated Confounder Selection]{Double Machine Learning and Automated Model Selection: A Cautionary Tale}

\date{This version: \today \\ First version: August 26, 2021}

\author[P.\ H\"{u}nermund, B.\ Louw, and I.\ Caspi]{Paul H\"{u}nermund$^{\dagger}$ \qquad
                Beyers Louw$^{\ddagger}$ \qquad Itamar Caspi$^{*}$}

\address{$^{\dagger}$Copenhagen Business School, Kilevej 14A, Frederiksberg, 2000, DK.}
\email{phu.si@cbs.dk}

\address{$^{\ddagger}$Maastricht University, Tongersestraat 53, 6211 LM Maastricht, NL.}
\email{jb.louw@maastrichtuniversity.nl}

\address{$^{*}$Bank of Israel, P.O.Box 780, 91007, Jerusalem, IL}
\email{itamar.caspi@boi.org.il}

\def\AmSTeX{$\cal A$\kern-.1667em\lower.5ex\hbox{$\cal M$}\kern-.125em
    $\cal S$-\TeX}
\def\BibTeX{{\rm B\kern-.05em{\sc i\kern-.025em b}\kern-.08em
    T\kern-.1667em\lower.7ex\hbox{E}\kern-.125emX}}

\begin{document}

    \begin{abstract}
		Double machine learning (DML) has become an increasingly popular tool for automated variable selection in high-dimensional settings. Even though the ability to deal with a large number of potential covariates can render selection-on-observables assumptions more plausible, there is at the same time a growing risk that endogenous variables are included, which would lead to the violation of conditional independence. This paper demonstrates that DML is very sensitive to the inclusion of only a few ``bad controls'' in the covariate space. The resulting bias varies with the nature of the theoretical causal model, which raises concerns about the feasibility of selecting control variables in a data-driven way.


    \end{abstract}

    \begin{flushright}
    \emph{``No causes in, no causes out."} \\
    --- Nancy Cartwright
    \end{flushright}
    

   \section{Introduction}

Machine learning approaches for selecting suitable control variables to establish causal identification in high-dimensional settings are gaining increasing attention (\citealp{Belloni2014, Chernozhukov2018}). Besides the evident benefits of automation for the analysis of high-dimensional data, this rising popularity can be explained by two specific advantages that applied researchers attribute to these methods. First, a mostly data-driven, automated procedure of model selection allows to systematize the research process and make it more transparent (\citealp{Athey2019}). And second, the ability to consider a large number of covariates --- possibly larger than the sample size --- could render selection-on-observables types of identification assumptions more plausible (\citealp{Belloni2014b}). For these reasons, automated variable selection has seen several recent applications in economics (\citealp{jones2019, Chang2020, Angrist2022}), finance (\citealp{Feng2020}), political science (\citealp{Dutt2018,Blackwell2021}), and organizational studies (\citealp{Vanneste2021}), as well as as the introduction of dedicated open source software libraries in \textit{R} and \textit{Python} (\citealp{Chernozhukov2019,Bach2022}).\footnote{See, for example, the vignette of the \textit{R}-package \textit{hdm}, which presents automated variable selection as a main application for illustrating the usefulness of double machine learning approaches.}

Double/debiased machine learning (DML) is a method developed to use regularized regression techniques, such as LASSO (\citealp{Tibshirani1996}) or $l_2$-boosting (\citealp{Buehlmann2003}), for variable selection in a high-dimensional causal inference setting (\citealp{Belloni2014b}). Compared to standard regularization on a single outcome equation, it seeks variables that are highly correlated with both treatment \emph{and} outcome, which immunizes the procedure against small approximation errors that inevitably arise when selecting among a large set of covariates. Consider the following system of partially linear equations
\begin{align}
    y &= \theta_0 d + g_0(x) + u, \label{eq:outcome} \\
    d &= m_0(x) + v, 
\end{align}
with primary interest in the causal effect $\theta_0$ of a treatment $D$ on outcome $Y$. The vector $X = (X_1, \hdots, X_p)$ consists of a set of covariates and $(U, V)$ are two disturbances with zero conditional mean. In settings where $X$ is high-dimensional and $g_0(\cdot)$ and $m_0(\cdot)$ are approximately linear and sparse, meaning that only a few elements of $X$ are important for predicting the treatment and outcome, regularization can be applied to automatically select the most suitable among a large set of potential control variables.

Yet, a na\"ive application of regularization to equation \ref{eq:outcome} can lead to substantial omitted variable bias (OVB), as it only selects variables that are highly correlated with the outcome $Y$, but not with the treatment $D$. The na\"ive approach therefore generally does not result in a root-$N$ consistent estimator for the structural parameter $\theta_0$ (\citealp{Chernozhukov2018}). Two main solutions to this problem are proposed in the literature: (a) partialling out, and (b) double selection, which both take into account the strength of association between $D$ and $X$. The former uses regularization to estimate the residuals of the outcome equation, $\rho^y = y - x' \pi_0^y$, and of the treatment equation,  $\rho^d = d - x' \pi_0^d$, with $\pi_0^y$ and $\pi_0^d$ being the respective coefficient vectors. It then finds the causal effect of interest $\hat{\theta}$ by regressing $\rho^y$ on $\rho^d$ (\citealp{Robinson1988}). The latter solution first determines suitable predictors for $Y$, then similarly finds predictors for $D$, and finally regresses $Y$ on the union of the selected controls. It can be shown that both approaches rely on doubly-robust moment conditions and are thus insensitive to approximation errors stemming from regularization (\citealp{Belloni2017,Chernozhukov2018}).

To causal inference scholars it is generally well known that model-free covariate selection is a theoretical impossibility --- a fact which was conceptualized by Pearl and Mackenzie under the rubric of the \emph{ladder of causation} (\citealp{Pearl2018}) and recently proven by \citet{Bareinboim2020}.  From this vantage point, the DML research program appears puzzling. If the starting point is a standard textbook regression equation, in which each variable $X_k$ is exogenous and the number of parameters $p$ is allowed to grow large, then variable selection is obviously feasible. Identification is achieved by assumption and the only task left for the machine learning algorithm is to pick the covariates with non-zero coefficients. But this ignores the problem that in reality not all covariates will be suitable controls.

The key identification assumption within the DML framework is \emph{ignorability} (\citealp{Imbens2004, Belloni2014}). Given the high-dimensional vector of control variables, treatment status is required to be conditionally independent of potential outcomes
\begin{equation}
    Y_{D=d} \independent D | X,
\end{equation}
with $Y_{D=d}$ denoting the potential outcome of $Y$ given treatment status $D=d$. This assumption can easily be violated, if $X$ includes variables that are not fully exogenous. In the following, we explore the consequences of violations of ignorability due to the presence of \emph{bad controls} in the conditioning set of the DML algorithm (\citealp{Angrist2009,Cinelli2022}). We focus on the LASSO case, which has received most attention so far (\citealp{jones2019, Vanneste2021,Angrist2022, Knaus2021}), presumably because of its appealing combination of interpretability and accuracy. However, as we will show, our arguments apply more broadly, also to the use of other machine learning algorithms for automated variable selection in a causal inference setting.

In a first step, we make precise the notion of bad controls in regression analyses by building on the \emph{backdoor criterion} from the graphical causal models literature (\citealp{Pearl1995,Cinelli2022}). We then show in simulations that DML is very sensitive to minor violations of the ignorability assumption. Depending on the exact source of endogeneity, the advantage of DML over na\"ive LASSO --- which was one of the main motivations for developing the method --- vanishes completely. This is because bad controls, although they do not necessarily exert a causal influence, are often highly correlated with the treatment or the outcome (since they are related to unobservables that affect $D$ or $Y$). Therefore, bad controls are very likely to be picked by DML, which has quantitative implications even if only a few endogenous variables are present in the conditioning set. We demonstrate this in an application of DML to the estimation of the gender wage gap using the data provided by \citet{Blau2017}. We find that the estimation results obtained by the original study differ in non-negligible ways compared to when marital status, which the literature identifies as being likely endogenous with respect to women’s labor-force decisions, is included in the covariate space.

Our study is related to a growing literature studying the performance of DML under various practically relevant data generating processes, of which most work has been focused on the omitted variable bias case. \citet{Wuethrich2021} show that double selection LASSO can exhibit substantial OVB as a result of variable under-selection in finite samples, even in favorable settings such as --- most relevant for this manuscript --- with uncorrelated, exogenous controls. Their findings render an application of the asymptotic distribution of $\sqrt{n}(\hat{\theta} - \theta_0)$ derived in \citet{Belloni2014} potentially problematic. Moreover, \citet{Chernozhukov2022} derive sharp bounds on the OVB in the presence of unobserved confounders, which can be used to perform sensitivity analysis. Instead, we focus on the case with endogenous, bad controls in the conditioning set.

Our results highlight significant pitfalls of automated, data-driven variable selection in high-dimensional settings. In particular, if numerous potential controls are considered, in an attempt to justify selection-on-observables, without theoretical background knowledge to guide the choice, the likelihood that some bad controls are accidentally included in the algorithm is high. Therefore, dealing with a large covariate space in an automated fashion might do little to approximate the ignorability assumption and is instead more useful to determine a suitable functional-form specification for a \emph{small} set of covariates, e.g., by considering higher-order polynomial terms (\citealp{Belloni2014, athey2017}). We show that this problem is not only prevalent for post-treatment variables or variables that are themselves considered outcome variables (\citealp{Vanneste2021}), so that researchers cannot rely on simple rules of thumb for variable inclusion. By contrast, each potential control requires its own careful identification argument based on domain knowledge\footnote{Early econometricians such as Tjalling Koopmans were of course well aware of this fact: ``Without resort to theory [...] conclusions relevant to the guidance of economic policies cannot be drawn'' (\citealp{Koopmans1947}).}, which is difficult to provide if the feature space is large and ultimately undermines the purpose of automated variable selection. We stress, however, that DML has broader applications, e.g., for the estimation of high-dimensional instrumental variable models (\citealp{Belloni2017}) and arbitrary do-calculus objects (\citealp{Jung2021}), as well as for data-splitting to reduce over-fitting. Our argument therefore specifically applies to the case when machine learning tools are used for the purpose of confounder selection.    
   \section{Preliminaries}	
\label{sec:theory}

An SCM is a 4-tuple $\langle V, U, F, P(u) \rangle$, where $V = \{V_1,\hdots,V_m\}$ is a set of endogenous variables that are determined in the model and $U$ denotes a set of (exogenous) background factors. $F$ is a set of functions $\{f_1,\hdots, f_m\}$ that assign values to the corresponding $V_i \in V$, such that $v_i \leftarrow f_i(pa_i, u_i)$, for $i = 1,\hdots, m$, and $PA_i \subseteq V \setminus V_i$.\footnote{The SCM literature uses assignment operators instead of equations to capture the asymmetric nature of causal relationships (\citealp{Huenermund2021}).} Finally, $P(u)$ is a probability function defined over the domain of $U$.

\begin{figure}[t]
\centering
	\subcaptionbox{\centering Good Control\label{fig1a}}[0.45\linewidth]{
		\begin{tikzpicture}[>=triangle 45, font=\footnotesize]
		\node[fill,circle,inner sep=0pt,minimum size=5pt,label={below:{D}}] (D) at (0,0) {};
		\node[fill,circle,inner sep=0pt,minimum size=5pt,label={above:{X}}] (X) at (2,1.5) {};
		\node[fill,circle,inner sep=0pt,minimum size=5pt,label={below:{Y}}] (Y) at (4,0) {};
		\draw[->,shorten >= 1pt] (X)--(D);
		\draw[->,shorten >= 1pt] (X)--(Y);
		\draw[->,shorten >= 1pt] (D)--(Y);
		\end{tikzpicture}
	}
	\subcaptionbox{\centering M-graph\label{fig1b}}[.45\linewidth]{
		\begin{tikzpicture}[>=triangle 45, font=\footnotesize]
		\node[fill,circle,inner sep=0pt,minimum size=5pt,label={below:{D}}] (D) at (0,0) {};
		\node[fill,circle,inner sep=0pt,minimum size=5pt,label={[yshift=0.5em]above:{X}}] (X) at (2,1) {};
		\node[fill,circle,inner sep=0pt,minimum size=5pt,label={below:{Y}}] (Y) at (4,0) {};
		\draw[->,shorten >= 1pt] (D)--(Y);
		\draw[<->,dashed,shorten >= 1pt, shorten <= 1pt] (X) to[bend right=60] (D);
		\draw[<->,dashed,shorten >= 1pt, shorten <= 1pt] (X) to[bend left=60] (Y);
		\end{tikzpicture}
	} \\
	\subcaptionbox{\centering Mediator\label{fig1c}}[.45\linewidth]{
		\begin{tikzpicture}[>=triangle 45, font=\footnotesize]
		\node[fill,circle,inner sep=0pt,minimum size=5pt,label={below:{D}}] (D) at (0,0) {};
		\node[fill,circle,inner sep=0pt,minimum size=5pt,label={above:{X}}] (X) at (2,1.5) {};
		\node[fill,circle,inner sep=0pt,minimum size=5pt,label={below:{Y}}] (Y) at (4,0) {};
		\draw[->,shorten >= 1pt] (D)--(X);
		\draw[->,shorten >= 1pt] (X)--(Y);
		\draw[->,shorten >= 1pt] (D)--(Y);
		\end{tikzpicture}
	}
	\subcaptionbox{\centering Confounded Mediator\label{fig1d}}[.45\linewidth]{
		\begin{tikzpicture}[>=triangle 45, font=\footnotesize]
		\node[fill,circle,inner sep=0pt,minimum size=5pt,label={below:{D}}] (D) at (0,0) {};
		\node[fill,circle,inner sep=0pt,minimum size=5pt,label={above:{X}}] (X) at (2,1.5) {};
		\node[fill,circle,inner sep=0pt,minimum size=5pt,label={below:{Y}}] (Y) at (4,0) {};
		\draw[->,shorten >= 1pt] (D)--(X);
		\draw[->,shorten >= 1pt] (X)--(Y);
		\draw[->,shorten >= 1pt] (D)--(Y);
		\draw[<->,dashed,shorten >= 1pt, shorten <= 1pt] (X) to[bend left=60] (Y);
		\end{tikzpicture}
	}
\captionsetup{justification=centering}
\caption{Directed acyclic graphs representing different structural causal models.}
\label{fig1}
\end{figure}
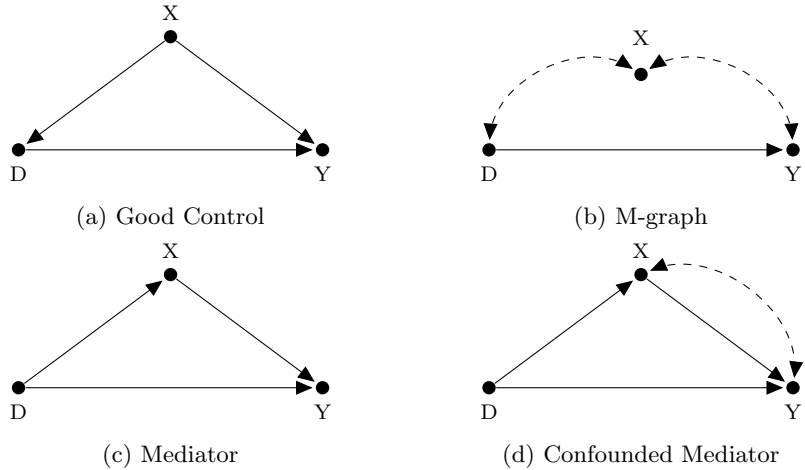

Every SCM defines a directed graph $\mathcal{G} \equiv (V, E)$, where $V$ is the set of endogenous variables, denoted as nodes (vertices) in the graph, and $E$ is a set of edges (links) pointing from $PA_i$ (the set of parent nodes) to $V_i$. An example is given by Fig.\ \ref{fig1a}, which corresponds to the SCM
\begin{equation}
\label{eq1}
\begin{aligned}
	x &\leftarrow f_1(u_1), \\
	d &\leftarrow f_2(x, u_2), \\
	y &\leftarrow f_3(d, x, u_3).
\end{aligned}
\end{equation}
Unobserved parents nodes induce a correlation between background factors in $U$. This is depicted by bidirected dashed arcs in the graph, which render the causal model \emph{semi-Markovian} (\citealp[pp.~30]{Pearl2009}). Fig.\ \ref{fig1b} depicts an example where the background factors of $X$ and $D$, as well as $X$ and $Y$ are correlated due to the presence of common influence factors that remain unobservable to the analyst. 

A sequence of edges connecting two nodes in $\mathcal{G}$ is called a \emph{path}. Paths can be either undirected or directed (i.e., following the direction of arrowheads). Since edges correspond to stimulus-response relations between variables in the underlying SCM (\citealp{Strotz1960}), directed paths represent the direction of causal influence in the graph. Due to the notion of causality being asymmetric (\citealp{Woodward2003,Cartwright2007}), directed cycles (i.e., loops from a node back to itself) are excluded, to rule out that a variable can be an (instantaneous) cause of itself. This assumption renders $\mathcal{G}$ acyclic.

A semi-Markovian causal graph $\mathcal{G}$ allows to decompose the distribution of the observed variables according to the factorization: $ P(v) = \sum_u \prod_i P(v_i|pa_i, u_i)P(u)$ (\citealp{Pearl2009}). The close connection between the topology of $\mathcal{G}$ and the probabilistic relationships --- in particular conditional independence relations --- between the variables that represent its nodes is further exemplified by the \emph{d-separation} criterion (\citealp{Pearl1988}). Consider three disjoint sets of variables, $X$, $Y$, and $Z$ in a DAG. These sets can either be connected via a (causal) chain, $X \rightarrow Z \rightarrow Y$, or a fork, $X \leftarrow Z \rightarrow Y$, where $Z$ acts as a common parent of $X$ and $Y$. A third possible configuration is the collider, $X \rightarrow Z \leftarrow Y$. In a chain and fork, conditioning on $Z$ renders $X$ and $Y$ conditionally independent, such that $X \independent Y | Z$. $Z$ is then said to ``d-separate'' or ``block the path between'' $X$ and $Y$. By contrast, in the collider structure, $X$ and $Y$ are independent from the outset, $X \independent Y | \emptyset$, whereas conditioning on $Z$ (or a descendant of $Z$; see (\citealp[def.\ 1.2.3]{Pearl2009}), would unblock the path, such that $X \not\independent Y | Z$.\footnote{Note that these d-separation relations hold for any distribution $P(v)$ over the variables in the model, in particular irrespective of any specific functional-form assumptions for $f_i$ and any distributional assumptions for $P(u)$ (\citealp{Huenermund2021}).}

D-separation gives rise to testable implications of graphical causal models (\citealp{Pearl2000}). Consider the following DAG:
\begin{center}
    \begin{tikzpicture}[>=triangle 45, font=\footnotesize]
    	\node[fill,circle,inner sep=0pt,minimum size=5pt,label={below:{X}}] (X) at (0,0) {};
    	\node[fill,circle,inner sep=0pt,minimum size=5pt,label={below:{Y}}] (Y) at (6,0) {};
    	\node[fill,circle,inner sep=0pt,minimum size=5pt,label={below:{Z}}] (Z) at (3,0) {};
    	\node[fill,circle,inner sep=0pt,minimum size=5pt,label={above:{W$_1$}}] (W1) at (1.5,1.5) {};
        \node[fill,circle,inner sep=0pt,minimum size=5pt,label={above:{W$_2$}}] (W2) at (4.5,1.5) {};
    	\draw[->,shorten >= 1pt, shorten <= 1pt] (X)--(Z) ;
    	\draw[->,shorten >= 1pt, shorten <= 1pt] (Z)--(Y);
    	\draw[->,shorten >= 1pt] (W1)--(X);
        \draw[->,shorten >= 1pt] (W1)--(Z);
    	\draw[->,shorten >= 1pt] (W2)--(Z);
        \draw[->,shorten >= 1pt] (W2)--(Y);
        \draw[->,shorten >= 1pt] (W2)--(Y);
        \draw[<->,dashed,shorten >= 1pt, shorten <= 1pt] (W1) to[bend left=45] (Z);
        \draw[<-,dashed,shorten >= 1pt, shorten <= 1pt] (W2) to[bend right=45] (Z);
    \end{tikzpicture}
\end{center}
This graph implies four d-separation relations between observed variables in the model: $W_1 \independent W_2$, $X \independent W_2$, $X \independent Y |W_2, Z$, and $Y \independent W_1 | W_2, Z$. They can be tested in the data with the help of a suitable conditional independence test, and if rejected, the hypothesized causal model can be discarded and refined.

Causal effects are defined in terms of interventions in the SCM, denoted by the $do(\cdot)$-operator (\citealp{Haavelmo1943,Strotz1960,Pearl1995}). For example, the intervention $do(D=d')$ in eq.\ \ref{eq1} entails to delete the function $f_2(\cdot)$, which normally assigns values to $D$, from the model and to replace it with the constant value $d'$. The target is then to estimate the post-intervention distribution of the outcome variable, $P(Y=y|do(D=d'))$, that results from this manipulation. Other quantities, such as the average causal effect (ACE) of a discrete change in treatment from $d'$ to $d''$, can then be computed by taking the difference in expected values: $E(Y|do(D=d'')) - E(Y|do(D=d'))$. However, since $P(y|do(d))$ is not directly observable in non-experimental data, it first needs to be transformed into a probability object that does not contain any do-operator before estimation can proceed (\citealp{Bareinboim2016,Huenermund2021}). This constitutes the \emph{identification} step in the graphical causal models literature (\citealp{Koopmans1950,Pearl2009}). 

\subsection{Backdoor Adjustment}

One popular strategy to identify the ACE is to control for confounding influence factors via covariate adjustment. This strategy can be rationalized with the help of the \emph{backdoor criterion} (\citealp{Pearl1995}).
\begin{definition}\label{def_backdoor}
	Given an ordered pair of treatment and outcome variables $(D,Y)$ in a causal graph $\mathcal{G}$, a set $X$ is backdoor admissible if it blocks (in the d-separation sense) every path between $D$ and $Y$ in the subgraph $\mathcal{G}_{\underline{D}}$, which is formed by deleting all edges from $\mathcal{G}$ that are emitted by $D$.
\end{definition}
\noindent Deleting edges emitted by $D$ from $\mathcal{G}$ ensures that all directed, causal paths between $D$ and $Y$ are kept open. The remaining paths are non-causal and thus create a spurious correlation between the treatment and outcome.\footnote{Since these paths point into $D$, they are said to ``enter through the backdoor''.} Consequently, a backdoor admissible set $X$ blocks all non-causal paths between $D$ and $Y$, while leaving the causal paths intact. The post-intervention distribution is then identifiable via the adjustment formula (\citealp{Pearl2009})
\begin{equation}
    P(y|do(d)) = \sum_{x} P(y|d,x)P(x).
\end{equation}
Since the right-hand side expression does not contain any do-operator, it can be estimated from observational data either by nonparametric methods, such as matching and inverse probability weighting, or, under additional functional-form assumptions, by parametric regression methods such as OLS.

However, following the d-separation criterion, correctly blocking backdoor paths via covariate adjustment can be intricate. Take Fig.\ \ref{fig1} as an example. In \ref{fig1a} there is one causal path, $D \rightarrow Y$, and one backdoor path, $D \leftarrow X \rightarrow Y$ (with $X$ being possibly vector-valued). Following the d-separation criterion, the backdoor path can be blocked by conditioning on $X$ so that only the causal influence of $D$ remains. By contrast, in the other depicted cases, controlling for $X$ would induce rather than reduce bias, thus, rendering $X$ a \emph{bad control} in these models. In Fig.\ \ref{fig1b}, which is known under the name of  \emph{m-graph} in the epidemiology literature (\citealp{Greenland2003}), $X$ exerts no causal influence on any variable in the graph. Still, there are unobserved confounders that result in a backdoor path, $D \dashleftarrow\dashrightarrow X \dashleftarrow\dashrightarrow Y$, which is already blocked however, since $X$ acts as a collider on this path. At the same time, since $X$ is a collider, conditioning on it (or any of its descendants) would unblock the path and therefore produce a spurious correlation. By contrast, $X$ does not lie on a backdoor path in Fig.\ \ref{fig1c}, but acts as a mediator between $D$ and $Y$. Controlling for $X$ would allow to filter out the direct effect of the treatment, $D \rightarrow Y$, from its mediated portion, $D \rightarrow X \rightarrow Y$, (\citealp{Imai2010}). However, this direct effect is generally different from the ACE, which has to be kept in mind for interpretation of results.\footnote{Additionally, following \citet{Imai2010}, identifying direct and indirect effects in a mediation setting requires the assumption of sequential ignorability, which is fulfilled in linear models with constant effects, but does not need to hold for every SCM.} Moreover, such an approach is risky, because if there are unobserved confounders between $X$ and $Y$, as depicted in Fig.\ \ref{fig1d}, $X$ becomes a collider on the path $D \rightarrow X \dashleftarrow\dashrightarrow Y$ and would thus lead to bias if conditioned on.\footnote{See \citet{Cinelli2022} for a more comprehensive discussion of bad controls in graphical causal models that goes beyond the scope of this paper.}

   \section{Simulation results}

In the following, we present a variety of simulations results to assess the magnitude of the bias introduced by including bad controls in the DML algorithm. We focus on the high-dimensional linear setting and apply double selection DML based on $l_1$-regularization to automatically select covariates. However, our argument is not specific to the LASSO case. In the online supplement, we present additional simulation results using $l_2$-boosting, which show very similar patterns. 

\begin{figure}[tp]
\centering
	\subcaptionbox{\centering Good Control\label{fig2a}}[.45\linewidth]{
		\begin{tikzpicture}[>=triangle 45, font=\footnotesize]
		\node[fill,circle,inner sep=0pt, draw=black, minimum size=5pt,label={below:{D}}] (D) at (0,0) {};
		\node[fill,circle,inner sep=0pt,minimum size=5pt,label={above:{X}}] (X) at (2,1.5) {};
		\node[fill,circle,inner sep=0pt,minimum size=5pt,label={below:{Y}}] (Y) at (4,0) {};
		\draw[->,shorten >= 1pt] (X)--(D) node [midway, sloped, above, minimum size=4pt] {0.8};
		\draw[->,shorten >= 1pt] (X)--(Y) node [midway, sloped, above, minimum size=4pt] {0.2};
		\draw[->,shorten >= 1pt] (D)--(Y) node [midway, sloped, below, minimum size=4pt] {1};
		\end{tikzpicture}
	}
	\subcaptionbox{\centering M-graph\label{fig2b}}[.45\linewidth]{
		\begin{tikzpicture}[>=triangle 45, font=\footnotesize]
		\node[fill,circle,inner sep=0pt,minimum size=5pt,label={below:{D}}] (D) at (0,0) {};
		\node[fill,circle,inner sep=0pt,minimum size=5pt,label={[yshift=0.2em]above:{X}}] (X) at (2,1) {};
		\node[fill,circle,inner sep=0pt,minimum size=5pt,label={below:{Y}}] (Y) at (4,0) {};
		\node[draw=black,fill=white,circle,inner sep=0pt,minimum size=5pt,label={above:{U$_1$}}] (U1) at (-0.2, 1.5) {};
		\node[draw=black,fill=white,circle,inner sep=0pt,minimum size=5pt,label={above:{U$_2$}}] (U2) at (4.2, 1.5) {};
		\draw[->,shorten >= 1pt] (D)--(Y) node [midway, sloped, below, minimum size=4pt] {1};
		\draw[->,shorten >= 1pt, shorten <= 1pt] (U1)--(D) node[midway, left] {$\sqrt{0.8}$};
		\draw[->,shorten >= 1pt, shorten <= 1pt] (U2)--(Y) node[midway, right] {$\sqrt{0.2}$};
		\draw[->,shorten >= 1pt, shorten <= 1pt] (U1)--(X) node[midway, sloped, above] {$\sqrt{0.8}$};
		\draw[->,shorten >= 1pt, shorten <= 1pt] (U2)--(X) node[midway, sloped, above] {$\sqrt{0.2}$};
		\end{tikzpicture}
	} \\
	\subcaptionbox{\centering Mediator\label{fig2c}}[.45\linewidth]{
		\begin{tikzpicture}[>=triangle 45, font=\footnotesize]
		\node[fill,circle,inner sep=0pt,minimum size=5pt,label={below:{D}}] (D) at (0,0) {};
		\node[fill,circle,inner sep=0pt,minimum size=5pt,label={above:{X}}] (X) at (2,1.5) {};
		\node[fill,circle,inner sep=0pt,minimum size=5pt,label={below:{Y}}] (Y) at (4,0) {};
		\draw[->,shorten >= 1pt] (D)--(X) node [midway, sloped, above, minimum size=4pt] {0.8};
		\draw[->,shorten >= 1pt] (X)--(Y) node [midway, sloped, above, minimum size=4pt] {0.2};
		\draw[->,shorten >= 1pt] (D)--(Y) node [midway, sloped, below, minimum size=4pt] {1};
		\end{tikzpicture}
	}
	\subcaptionbox{\centering Confounded Mediator\label{fig2d}}[.45\linewidth]{
		\begin{tikzpicture}[>=triangle 45, font=\footnotesize]
		\node[fill,circle,inner sep=0pt,minimum size=5pt,label={below:{D}}] (D) at (0,0) {};
		\node[fill,circle,inner sep=0pt,minimum size=5pt,label={above:{X}}] (X) at (2,1) {};
		\node[fill,circle,inner sep=0pt,minimum size=5pt,label={below:{Y}}] (Y) at (4,0) {};
		\draw[->,shorten >= 1pt] (D)--(X) node [midway, sloped, above, minimum size=4pt] {0.8};
		\draw[->,shorten >= 1pt] (X)--(Y) node [midway, sloped, above, minimum size=4pt] {0.2};
		\draw[->,shorten >= 1pt] (D)--(Y) node [midway, sloped, below, minimum size=4pt] {1};
		\node[draw=black,fill=white,circle,inner sep=0pt,minimum size=5pt,label={above:{U$_1$}}] (U1) at (4.2, 1.5) {};
		\draw[->,shorten >= 1pt, shorten <= 1pt] (U1)--(X) node[midway, sloped, above] {$\sqrt{0.2}$};
		\draw[->,shorten >= 1pt, shorten <= 1pt] (U1)--(Y) node[midway, right] {$\sqrt{0.2}$};
		\end{tikzpicture}
	}
\captionsetup{justification=centering}
\caption{Baseline parametrization of the simulations.}
\label{fig2}
\end{figure}
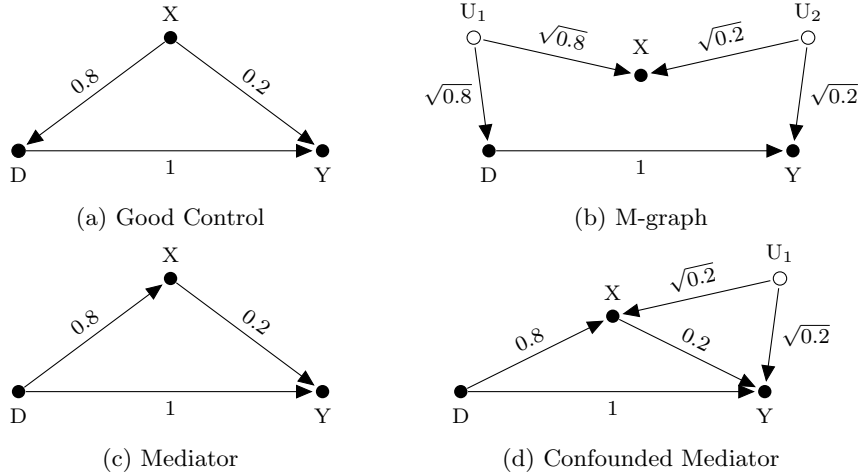

Since DML is specifically designed to spot variables that are mainly correlated with the treatment, which is the reason for its superior performance compared to na\"ive LASSO, for our baseline specification, we set a higher correlation between the controls and the treatment than with the outcome. We fix the sample size at $n=1,000$ and number of covariates at $p=100$. To introduce sparsity, only $q=10$ out of these variables are specified as having non-zero coefficients. The treatment effect $\theta_0$ is constant and set equal to one. All exogenous nodes (which do not receive any incoming arrows) are specified as standard normal. In the baseline, parameters are chosen in such a way that the strength, measured as the product of structural coefficients, of each path connecting the (non-zero) covariates and the treatment is equal to $b_1 = 0.8$. Similarly, the strength of paths connecting the covariates and the outcome is set to  $b_2 = 0.2$ (Fig.\ \ref{fig2} depicts the baseline parametrization in form of a path diagram with associated coefficients as edge labels, hollow circles indicate unobserved variables). 

Following the double selection method, we then regress $Y$ on $X$ using LASSO and record the variables with estimated non-zero coefficients. We do the same for a LASSO regression of $D$ on $X$. Finally, we regress $Y$ on the union of variables in $X$ that have been picked in the preceding two LASSO regressions, this time using standard OLS. We record the estimated coefficients for the treatment effect of interest $\hat{\theta}$ across 10,000 simulation runs. In addition, we compare double selection with the na\"ive (post)LASSO method, in which we repeat the previous protocol but without the second step of regressing $D$ on $X$. I.e., in na\"ive LASSO variables are only selected once for the outcome regression, disregarding their correlation with the treatment. To summarize the estimation algorithms:

\bigskip

\begin{itemize}
    \item \textbf{DML (double selection)}
        \begin{enumerate}
            \item Regress $Y$ on $X$ via LASSO and record all $X_k$ with nonzero coefficients
            \item Regress $D$ on $X$ via LASSO and record all $X_{k'}$ with nonzero coefficients
            \item Regress $Y$ on the union of all $X_k$ and $X_{k'}$
        \end{enumerate}
    \item \textbf{Na\"ive (post-)LASSO}
        \begin{enumerate}
            \item Regress $Y$ on $X$ via LASSO and record all $X_k$ with nonzero coeffcients
            \item Regress $Y$ on all $X_k$
        \end{enumerate}
\end{itemize}

\begin{figure}[pt]
\centering
    \begin{tabular}{ccc}
     \underline{Double Machine Learning} & \underline{Na\"ive Lasso} \\ \addlinespace
     \multicolumn{2}{l}{(a) Good Control} \\
     \includegraphics[width=60mm]{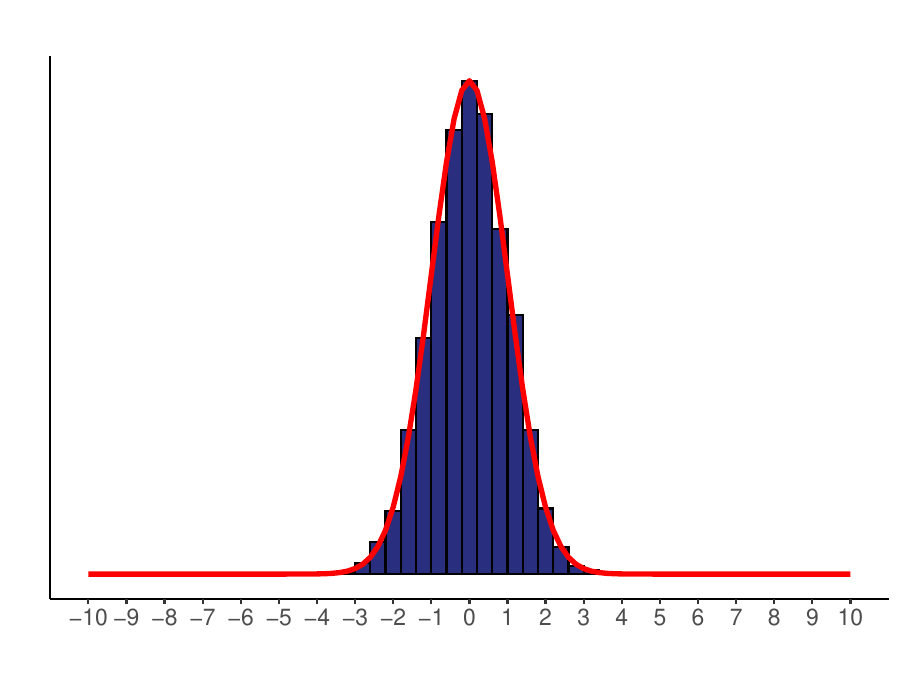} & \includegraphics[width=60mm]{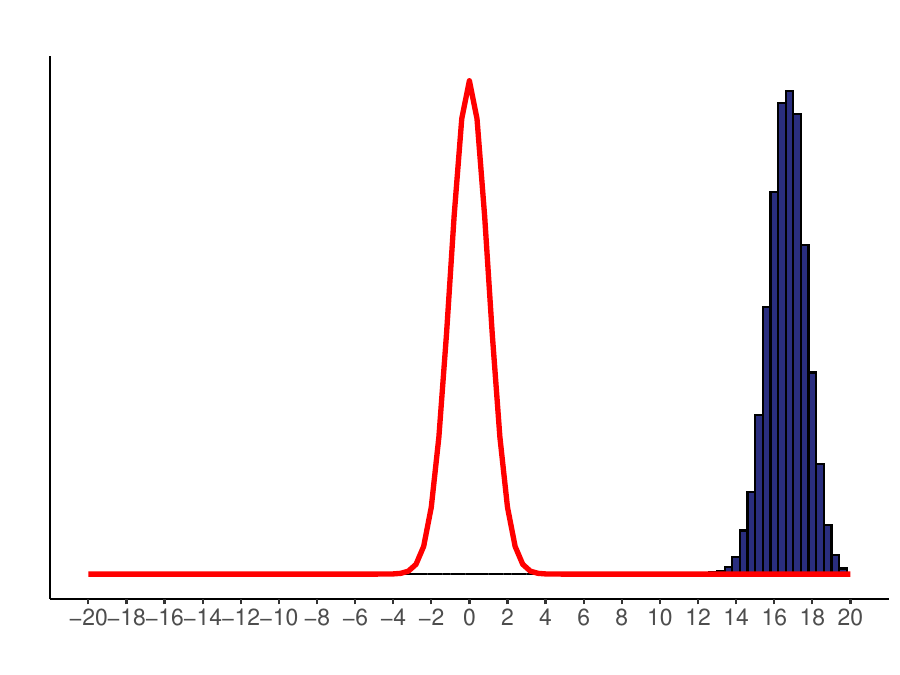} \\
     \multicolumn{2}{l}{(b) M-graph} \\
     \includegraphics[width=60mm]{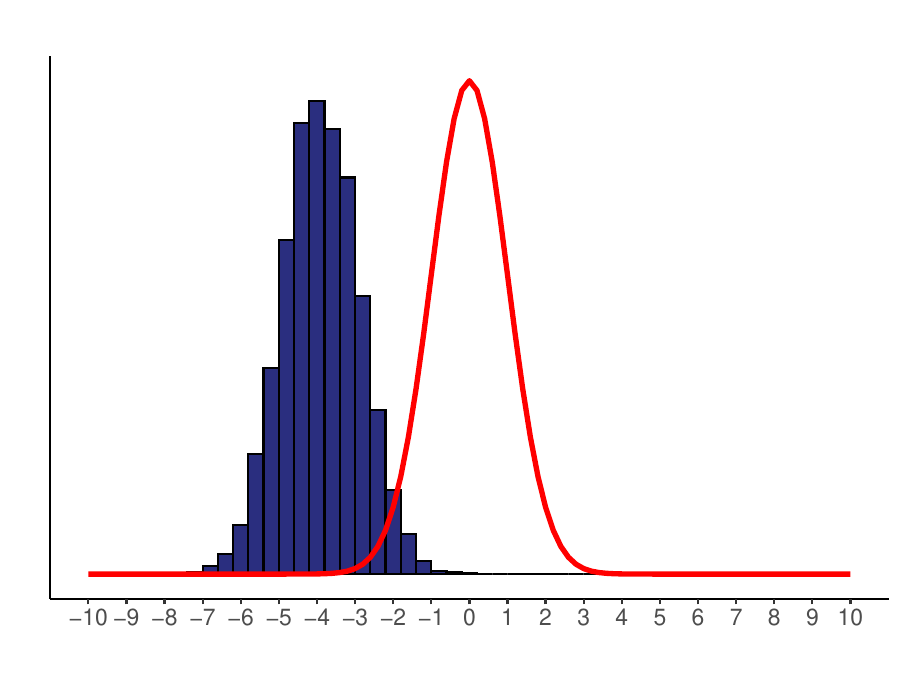} & \includegraphics[width=60mm]{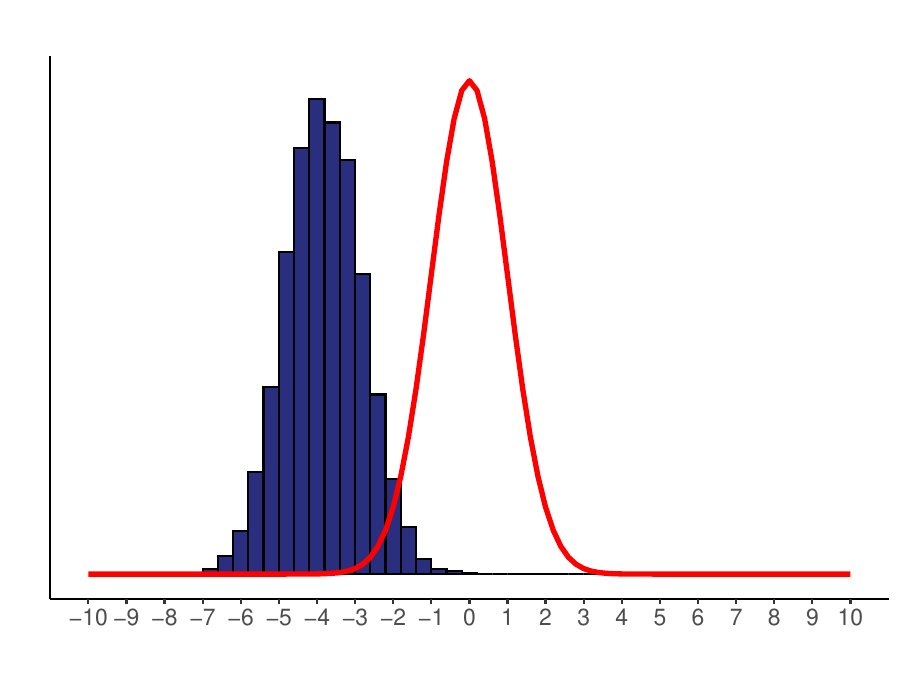} \\
     \multicolumn{2}{l}{(c) Mediator (direct effect)} \\
     \includegraphics[width=60mm]{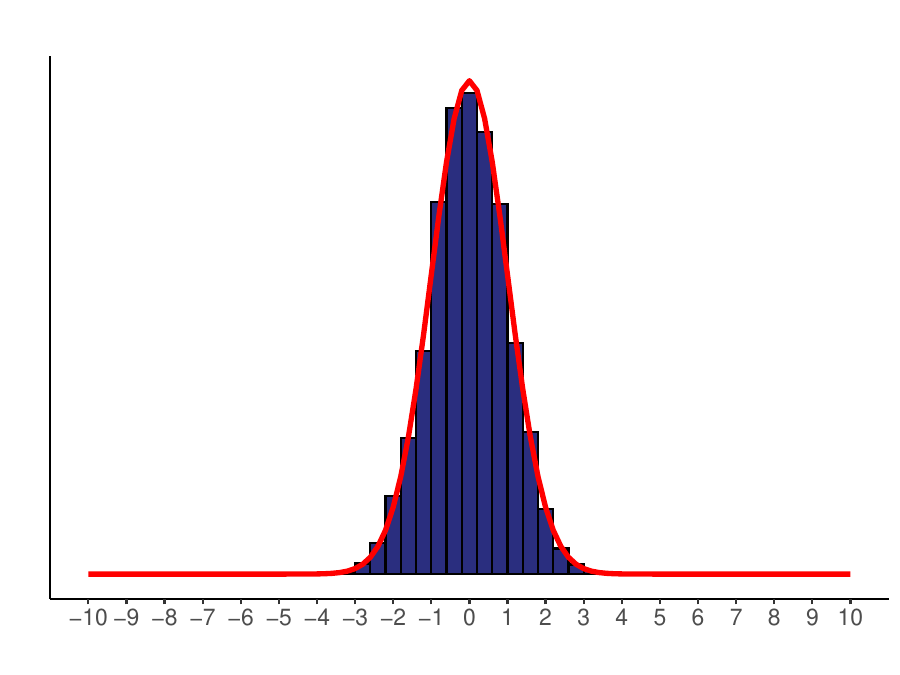} & \includegraphics[width=60mm]{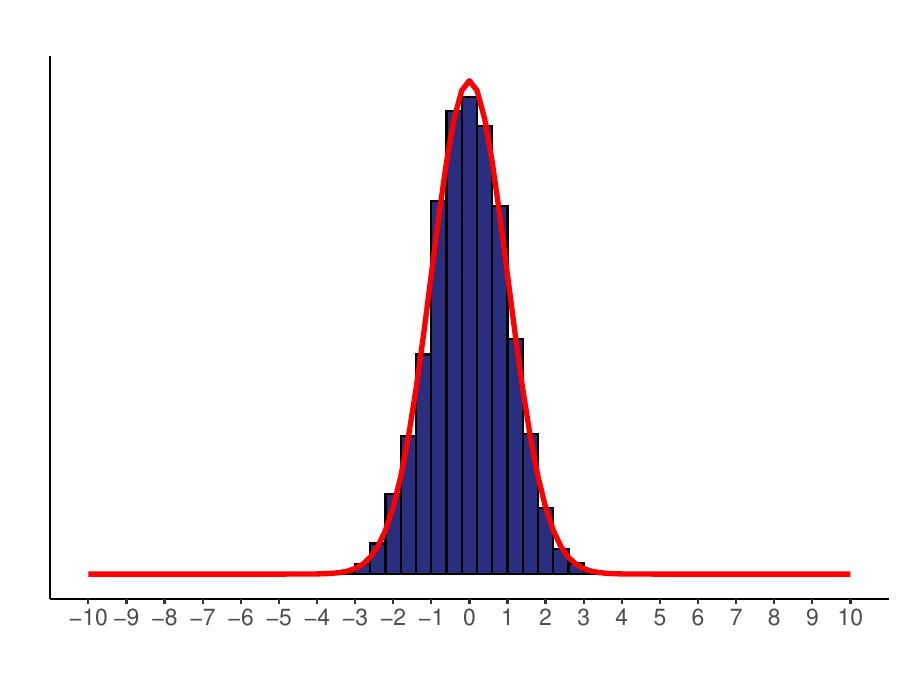} \\
     \multicolumn{2}{l}{(d) Confounded Mediator (direct effect)} \\
     \includegraphics[width=60mm]{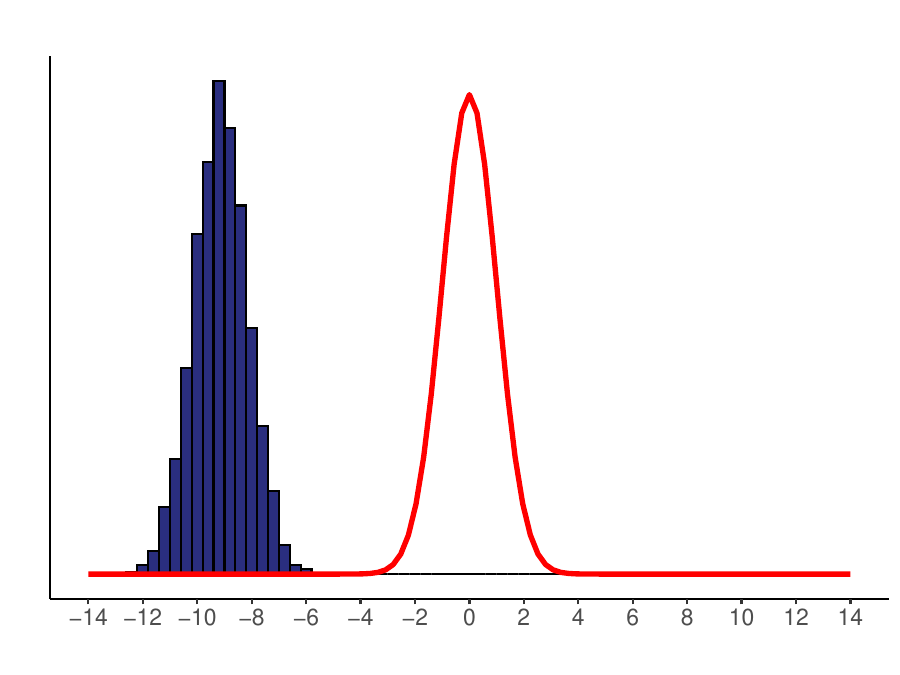} & \includegraphics[width=60mm]{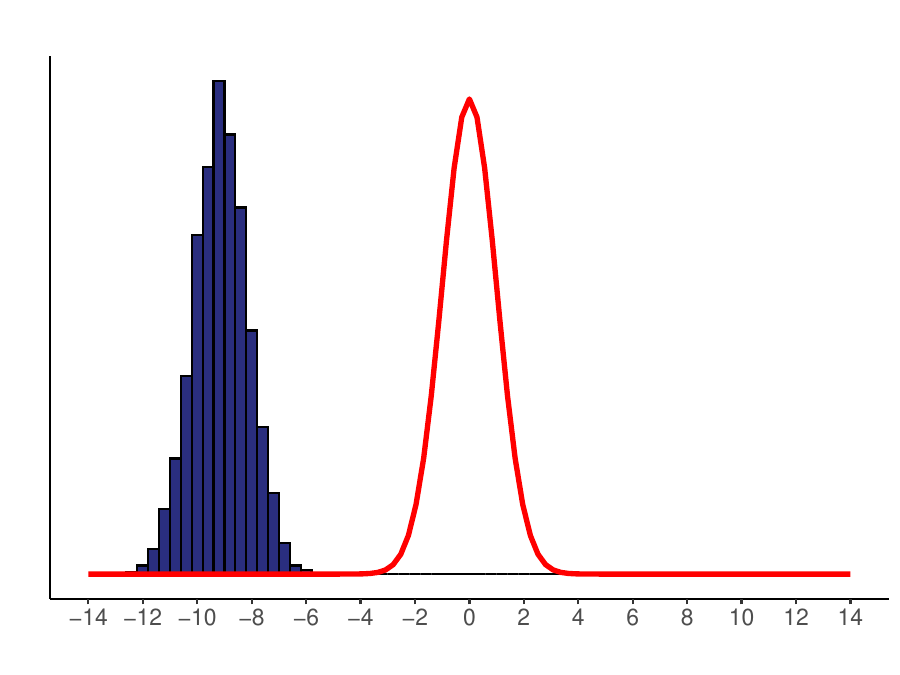} \\
    \end{tabular}
\captionsetup{justification=centering}
\caption{Performance of DML compared to na\"ive LASSO for different causal models.}
\label{fig3}
\end{figure}

Fig.\ \ref{fig3} shows simulation results using centered and studentized quantities, next to their theoretical (standard normal) distribution. In panel (a) we observe the familiar picture from \citet{Belloni2014}. DML is able to reliably filter out the good controls from irrelevant covariates, which leads to a distribution that closely matches the theoretical one. By contrast, na\"ive LASSO fails to pick relevant control variables that are only weakly correlated with the outcome, translating into substantial bias. However, this result reverses for the m-graph in panel (b). Here, the covariates are bad controls, due to the collider structure, and should not be included in the regression. They are nonetheless highly correlated with the treatment and thus get picked by the DML, leading to biased causal effect estimates. In fact, the advantage that DML had over na\"ive LASSO in (a) vanishes completely ($bias^{DML} = -0.120$, and $bias^{LASSO} = -0.119$). Interestingly, given the chosen parameterization with only a moderately high correlation between the covariates and the outcome, the na\"ive approach consistently selects fewer bad controls than DML. The mode of the number of controls selected across simulations is 5 for the na\"ive LASSO and 10 for DML.

In panel (c) we investigate the mediator case. Now, the covariates are post-treatment variables, which nonetheless end up getting selected as controls by both the na\"ive LASSO and DML. According to the discussion in Sec.\ \ref{sec:theory}, this allows to consistently estimate the direct effect of the treatment. However, the researcher needs to keep this change of target parameter in mind for interpretation, since both  na\"ive LASSO and DML are unable to consistently estimate the total effect of treatment. Moreover, once we introduce a confounded mediator in panel (d), both DML and  na\"ive LASSO perform equally poorly. The direct effect cannot be consistently estimated in this model, as neither controlling for the mediators nor leaving them out would be sufficient for identification. The total effect of treatment is likewise not estimable via DML (but would be by a simple regression of $Y$ on $D$).

\begin{table}[t]
\captionsetup{justification=centering}
\caption{Bias obtained from DML under various parameter constellations ($\theta_0 = 1$).}
\label{tab:vary_parameters}
\centering
    \begin{tabular}{lccccc}
    \addlinespace
    \toprule
    \multicolumn{1}{c}{$(b_1, b_2)=$} & $(0.8, 0.2)$ & $(0.6, 0.4)$ & $(0.5, 0.5)$ & $(0.4, 0.6)$ & $(0.2, 0.8)$ \\ 
    \midrule
    Good Control & 0.000 & 0.000 & 0.000 & 0.000 & 0.000 \\
    M-graph & -0.120 & -0.172 & -0.179 & -0.174 & -0.126 \\
    Mediator & -0.001 & -0.001 & -0.001 & 0.000 & 0.000 \\
    Confounded Mediator & -0.534 & -0.480 & -0.417 & -0.343 & -0.178\\
    \addlinespace \midrule
    \multicolumn{1}{c}{$q=$} & $1$ & $5$ & $10$ & $20$ & $50$ \\
    \midrule
    Good Control & 0.000 & 0.000 & 0.000 & 0.000 & 0.000 \\
    M-graph & -0.054 & -0.105 & -0.120 & -0.128 & -0.134 \\
    Mediator & 0.000 & -0.001 & -0.001 & -0.001 & -0.001 \\
    Confounded Mediator & -0.134 & -0.401 & -0.534 & -0.641 & -0.728 \\
        \addlinespace
    \bottomrule
    \end{tabular}
\end{table}

Table \ref{tab:vary_parameters} depicts the bias obtained from DML for varying parameter constellations. In the top panel, we study performance depending on whether there is a higher strength of association between the covariates and the treatment or the outcome. For the two bad control cases, i.e., the m-graph and confounded mediator, substantial bias arises regardless of the chosen parametrization. When taking into account the change of target parameter from the total to the direct effect, bias is low for the simple mediator model across all setups. Moreover, the DML generally performs well in the good control case, although bias becomes slightly larger when the strength of association is stronger with the outcome than the treatment (see also the $n=100$ case in the supplemental material).

In the bottom panel of Table \ref{tab:vary_parameters}, we vary the number of covariates with non-zero coefficients $q$ (with $b_1=0.8, b_2=0.2$, as before), while the total number of variables considered in the conditioning set remains fixed at $p=100$.\footnote{In unreported analyses, we find similar results if bad controls are mixed with good controls instead of irrelevant (zero-coefficient) ones. The two cases are conceptually similar since the DML either picks the good or leaves out the irrelevant controls, resulting in a zero bias baseline, which then gets distorted by the selected bad controls.} Interestingly, a noticeable bias (around 5 percent for the m-graph and 13 percent for the confounded mediator) arises already with one bad control out of a hundred, and increases monotonically in $q$. The bias for the direct effect remains low for the simple mediator model. 

In the online supplement we present additional simulations with varying $p$ and $n$. In particular, we explore the case of $n=100$ since DML is often proposed as a technique for dealing with a large number of predictors in relatively small data sets ($p \gg n$). We find results that are in line with the ones presented here in the main text.
   \section{Application}

\begin{figure}[t]
\centering
		\begin{tikzpicture}[>=triangle 45, font=\footnotesize]
		\node[fill,circle,inner sep=0pt,minimum size=5pt,label={left:{Gender}}] (Gender) at (0,0) {};
		\node[fill,circle,inner sep=0pt,minimum size=5pt,label={above:{Marital Status}}] (Marital Status) at (2,1.5) {};
		\node[fill,circle,inner sep=0pt,minimum size=5pt,label={below:{X}}] (X) at (2,-1) {};
		\node[fill,circle,inner sep=0pt,minimum size=5pt,label={right:{Wages}}] (Wages) at (4,0) {};
		\draw[->,shorten >= 1pt] (Gender)--(Marital Status);
		\draw[->,shorten >= 1pt] (Marital Status)--(Wages);
		\draw[->,shorten >= 1pt] (Gender)--(Wages);
		\node[draw=black,fill=white,circle,inner sep=0pt,minimum size=5pt,label={above:{U}}] (U) at (4.2, 1.5) {};
		\draw[->,shorten >= 1pt, shorten <= 1pt] (U)--(Marital Status) ;
		\draw[->,shorten >= 1pt, shorten <= 1pt] (U)--(Wages);
		\draw[->,shorten >= 1pt] (X)--(Gender);
		\draw[->,shorten >= 1pt] (X)--(Wages);
		\end{tikzpicture}
\captionsetup{justification=centering}
\caption{Causal diagram for the gender wage gap study in \citet{Blau2017}.}
\label{fig4}
\end{figure}
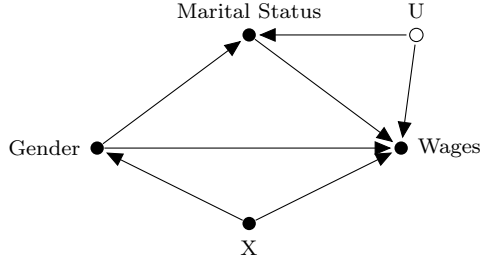

For an application to real-world data, we make use of the \emph{Panel Study of Income Dynamics} (PSID) microdata provided by \citet{Blau2017}. They estimate the extent of the gender wage gap in six waves of the PSID between 1981 and 2011. For their full specification, they employ a rich set of 50 control variables (as described in Section IV of their online appendix), including individual-level information on education, experience, race, occupation, unionization, as well as regional and industry characteristics. However, Blau and Kahn deliberately decide to exclude marital status and number of children from their regressions, because these variables ``are likely to be endogenous with respect to women’s labor-force decisions'' (p.\ 797). Although the source of this endogeneity is not further discussed, we find it plausible that marital status acts as a confounded mediator, since it is likely influenced by the same unobserved background factors that also affect wages (see Fig.\ \ref{fig4}).

From the PSID data, we can infer a woman's marital status based on whether she is recorded as ``legally married wife'' in her relation to the household head (men are by default indicated as household heads). Our goal is to test the sensitivity of the estimated (adjusted) gender wage gap to the inclusion of this potentially bad control. As a benchmark, we regress log wages on a female dummy and the original set of controls for each wave separately. We then employ DML using the double selection method, which allows us to include all interactions of the control variables up to degree 2. In a last step, we add marital status and its interactions to the model matrix in the DML. 

Results are shown in Table \ref{tab:application}. The estimated gender wage gaps in the OLS specifications range from $(1-\exp(-0.249)) \approx 22$ percentage points in 1981 to approximately $13.5$ p.p.\ in 2011. Most of the convergence between male and female wages happens in the 1980s, which coincides with the results in \citet{Blau2017}. Although the DML relies on a much larger set of covariates, the results are very similar to OLS. We find greater discrepancies, however, when marital status is included in the feature space. Across all six waves, marital status (as well as several interactions) ends up getting picked as control by the double selection DML. This has non-negligible impact on the estimated gender wage gaps, which are 10.6\% larger on average, in absolute terms, compared to the benchmark OLS. Under the assumption that marital status is a confounded mediator, larger gaps might be the result of a negative correlation between wages and the decision to get married, induced by unobservables. The respective path gets activated when marital status, as a collider, is conditioned on. Thus, the example demonstrates how having only one endogenous control within a large covariate space, paired with a flexible DML approach, can substantially affect the quantitative conclusions drawn from a study.\footnote{We find even larger differences if marital status is included as a single regressor, without interacting it with other covariates; see Table S4 in the online supplemental material.}

\begin{table}[t]
\captionsetup{justification=centering}
\caption{Effect of gender on log wages using PSID data from \citet{Blau2017} (standard errors in parentheses).}
\label{tab:application}
\centering
    \begin{tabular}{lcccccc}
    \addlinespace
    \toprule
    \multicolumn{1}{c}{Wave =} & 1981 & 1990 & 1999 & 2007 & 2009 & 2011 \\
    \midrule
    OLS & -0.249 & -0.137 & -0.158 & -0.168 & -0.157 & -0.145 \\
    & (0.016) & (0.014) & (0.016) & (0.015) & (0.015) & (0.016) \\
    DML & -0.268 & -0.139 & -0.158 & -0.164 & -0.157 & -0.136 \\
    & (0.017) & (0.015) & (0.016) & (0.016) & (0.016) & (0.017) \\
    DML incl. & -0.270 & -0.154 & -0.173 & -0.190 & -0.179 & -0.163 \\
    \ \emph{marital status} & (0.022) & (0.019) & (0.020) & (0.019) & (0.020) & (0.021) \\
    \addlinespace
    \bottomrule
    \end{tabular}
\end{table}

   \section{Discussion}

In this paper, we demonstrate the sensitivity of automated confounder selection using double machine learning approaches to the inclusion of bad controls in the conditioning set. In our simulations, only when covariates are strictly exogenous, DML shows superior performance to na\"ive LASSO. In all other cases it performs equally poorly or worse. Furthermore, our empirical application illustrates that a non-negligible bias can already occur with a small number of endogenous variables in an otherwise much larger covariate space.

These results highlight why it may be problematic to use machine learning techniques for the automatic selection of control variables in regression settings. While the ability to deal with a large set of potential controls in an automated fashion can add to the plausibility of selection-on-observable assumptions, there is an increasing chance that bad controls might be included unintentionally if the covariate space grows large. Automated approaches thus turn out to be a double-edged sword, in particular if the number of control variables becomes so large that the researcher is unable to provide a sufficient theoretical discussion for each of them. We show that simple rules of thumb, such as restricting the conditioning set to only pre-treatment variables, do not offer adequate safeguards against this problem. Indeed, as Figure \ref{fig1b} shows, our results are not limited to post-treatment variables. The intricacies of the backdoor criterion (recall, e.g., the implications of subtle differences between Figures \ref{fig1c} and \ref{fig1d}) imply that a vague intuition, without the the guidance of a causal model, will likely be insufficient to ensure causal identification.

Because DML already assumes unconfounded covariates (\citealp[sec.\ 5]{Chernozhukov2018}), using its ability to handle a large feature space in order to justify unconfoundedness, ultimately leads to a circular argument. As long as causal inference is the goal, the analyst needs to provide a theoretical justification for the exogeneity of each of the considered control variables individually, which echoes \citet{Cartwright1989}'s familiar adage: ``no causes in, no causes out.'' Since this is difficult to achieve in high-dimensional settings, from a practical standpoint, smaller models that focus only on the most relevant covariates for a given context might actually be preferable. 

For the purpose of automated model selection, causal discovery algorithms from the artificial intelligence literature could represent a viable alternative (\citealp{Spirtes2000, Peters2017}). These methods do not rely on unconfoundedness and clarify the possibilities for data-driven causal learning based on a minimal set of assumptions. A key insight from this literature is that causal structures can only be learned up to a certain equivalence class from data. As a result, the ultimate justification for a particular causal model needs to come from theoretical background knowledge (\citealp{Bareinboim2020}). The same applies to DML, which is a highly effective tool, e.g., for selecting suitable functional specifications involving a small set of controls in a data-driven way. In big data settings with a large number of potential covariates, however, DML needs to be applied carefully to avoid bad controls and to ensure robust results.

	
	\section*{Acknowledgements}
	The authors are grateful to Elias Bareinboim, Victor Chernozhukov, Jevgenij Gamper, Daniel Millimet, Judea Pearl, and seminar participants at Booking.com, Microsoft, RWTH Aachen, and Vinted for useful comments and suggestions.
    

    \bibliography{dml_references.bib}

\begin{thebibliography}{}

\bibitem[\protect\citeauthoryear{Angrist and Frandsen}{Angrist and
  Frandsen}{2022}]{Angrist2022}
Angrist, J.~D. and B.~Frandsen (2022).
\newblock Machine labor.
\newblock {\em J Labor Econ\/}~{\em 40\/}(S1), S97--S140.

\bibitem[\protect\citeauthoryear{Angrist and Pischke}{Angrist and
  Pischke}{2009}]{Angrist2009}
Angrist, J.~D. and J.-S. Pischke (2009).
\newblock {\em {M}ostly {H}armless {E}conometrics: {A}n {E}mpiricist's
  {C}ompanion}.
\newblock Princeton University Press.

\bibitem[\protect\citeauthoryear{Athey}{Athey}{2019}]{Athey2019}
Athey, S. (2019).
\newblock The impact of machine learning on economics.
\newblock In A.~Agrawal, J.~Gans, and A.~Goldfarb (Eds.), {\em The Economics of
  Artificial Intelligence: An Agenda}. Chicago, IL, USA: The University of
  Chicago Press.

\bibitem[\protect\citeauthoryear{Athey and Imbens}{Athey and
  Imbens}{2017}]{athey2017}
Athey, S. and G.~W. Imbens (2017).
\newblock The state of applied econometrics: Causality and policy evaluation.
\newblock {\em J Econ Perspect\/}~{\em 31\/}(2), 3--32.

\bibitem[\protect\citeauthoryear{Bach, Chernozhukov, Kurz, and Spindler}{Bach
  et~al.}{2022}]{Bach2022}
Bach, P., V.~Chernozhukov, M.~S. Kurz, and M.~Spindler (2022).
\newblock Doubleml - an object-oriented implementation of double machine
  learning in python.
\newblock {\em J Mach Learn Res\/}~{\em 23\/}(53), 1--6.

\bibitem[\protect\citeauthoryear{Bareinboim, Correa, Ibeling, and
  Icard}{Bareinboim et~al.}{2022}]{Bareinboim2020}
Bareinboim, E., J.~D. Correa, D.~Ibeling, and T.~Icard (2022, February).
\newblock On pearl’s hierarchy and the foundations of causal inference.
\newblock {\em Probabilistic and Causal Inference: The Works of Judea Pearl\/},
  507--556.

\bibitem[\protect\citeauthoryear{Bareinboim and Pearl}{Bareinboim and
  Pearl}{2016}]{Bareinboim2016}
Bareinboim, E. and J.~Pearl (2016).
\newblock Causal inference and the data-fusion problem.
\newblock {\em Proc Natl Acad Sci\/}~{\em 113}, 7345--7352.

\bibitem[\protect\citeauthoryear{Belloni, Chernozhukov, Fern\'andez-Val, and
  Hansen}{Belloni et~al.}{2017}]{Belloni2017}
Belloni, A., V.~Chernozhukov, I.~Fern\'andez-Val, and C.~Hansen (2017).
\newblock Program evaluation and causal inference with high-dimensional data.
\newblock {\em Econometrica\/}~{\em 85}, 233--298.

\bibitem[\protect\citeauthoryear{Belloni, Chernozhukov, and Hansen}{Belloni
  et~al.}{2014a}]{Belloni2014b}
Belloni, A., V.~Chernozhukov, and C.~Hansen (2014a).
\newblock High-dimensional methods and inference on structural and treatment
  effects.
\newblock {\em J Econ Perspect\/}~{\em 28\/}(2), 29--50.

\bibitem[\protect\citeauthoryear{Belloni, Chernozhukov, and Hansen}{Belloni
  et~al.}{2014b}]{Belloni2014}
Belloni, A., V.~Chernozhukov, and C.~Hansen (2014b).
\newblock Inference on treatment effects after selection among high-dimensional
  controls.
\newblock {\em Rev Econ Stud\/}~{\em 81}, 608--650.

\bibitem[\protect\citeauthoryear{Blackwell and Olson}{Blackwell and
  Olson}{2021}]{Blackwell2021}
Blackwell, M. and M.~P. Olson (2021).
\newblock Reducing model misspecification and bias in the estimation of
  interactions.
\newblock {\em Polit Anal\/}~{\em 30\/}(4), 495--514.

\bibitem[\protect\citeauthoryear{Blau and Kahn}{Blau and Kahn}{2017}]{Blau2017}
Blau, F.~D. and L.~M. Kahn (2017).
\newblock The gender wage gap: Extent, trends, and explanations.
\newblock {\em J Econ Lit\/}~{\em 55}, 789--865.

\bibitem[\protect\citeauthoryear{B\"{u}hlmann and Yu}{B\"{u}hlmann and
  Yu}{2003}]{Buehlmann2003}
B\"{u}hlmann, P. and B.~Yu (2003).
\newblock Boosting with the l\textsubscript{2} loss: Regression and
  classification.
\newblock {\em J Am Stat Assoc\/}~{\em 98}, 324--339.

\bibitem[\protect\citeauthoryear{Cartwright}{Cartwright}{1989}]{Cartwright1989}
Cartwright, N. (1989).
\newblock {\em Nature's Capacities and Their Measurement}.
\newblock Oxford, UK: Clarendon Press.

\bibitem[\protect\citeauthoryear{Cartwright}{Cartwright}{2007}]{Cartwright2007}
Cartwright, N. (2007).
\newblock {\em Hunting Causes and Using Them}.
\newblock Cambridge, UK: Cambridge University Press.

\bibitem[\protect\citeauthoryear{Chang}{Chang}{2020}]{Chang2020}
Chang, N.-C. (2020).
\newblock Double/debiased machine learning for difference-in-differences
  models.
\newblock {\em Econom J\/}~{\em 23\/}(2), 177--191.

\bibitem[\protect\citeauthoryear{Chernozhukov, Chetverikov, Demirer, Duflo,
  Hansen, Newey, and Robins}{Chernozhukov et~al.}{2018}]{Chernozhukov2018}
Chernozhukov, V., D.~Chetverikov, M.~Demirer, E.~Duflo, C.~Hansen, W.~Newey,
  and J.~Robins (2018).
\newblock Double/debiased machine learning for treatment and structural
  parameters.
\newblock {\em Econom J\/}~{\em 21}, C1--C68.

\bibitem[\protect\citeauthoryear{Chernozhukov, Cinelli, Newey, Sharma, and
  Syrgkanis}{Chernozhukov et~al.}{2022}]{Chernozhukov2022}
Chernozhukov, V., C.~Cinelli, W.~Newey, A.~Sharma, and V.~Syrgkanis (2022).
\newblock Long story short: Omitted variable bias in causal machine learning.
\newblock \url{https://doi.org/10.48550/arXiv.2112.13398}.

\bibitem[\protect\citeauthoryear{Chernozhukov, Hansen, and
  Spindler}{Chernozhukov et~al.}{2019}]{Chernozhukov2019}
Chernozhukov, V., C.~Hansen, and M.~Spindler (2019).
\newblock {\em High-dimensional Metrics in R}.

\bibitem[\protect\citeauthoryear{Cinelli, Forney, and Pearl}{Cinelli
  et~al.}{2022}]{Cinelli2022}
Cinelli, C., A.~Forney, and J.~Pearl (2022).
\newblock A crash course in good and bad controls.
\newblock {\em Sociol Method Res\/}.
\newblock forthcoming.

\bibitem[\protect\citeauthoryear{Dutt and Tsetlin}{Dutt and
  Tsetlin}{2018}]{Dutt2018}
Dutt, P. and I.~Tsetlin (2018).
\newblock Income distribution and economic development: Insights from machine
  learning.
\newblock {\em Econ Polit\/}~{\em 33\/}(1), 1--36.

\bibitem[\protect\citeauthoryear{Feng, Giglio, and Xiu}{Feng
  et~al.}{2020}]{Feng2020}
Feng, G., S.~Giglio, and D.~Xiu (2020).
\newblock Taming the factor zoo: A test of new factors.
\newblock {\em J Financ\/}~{\em 75\/}(3), 1327--1370.

\bibitem[\protect\citeauthoryear{Greenland}{Greenland}{2003}]{Greenland2003}
Greenland, S. (2003).
\newblock Quantifying biases in causal models: Classical confounding vs
  collider-stratification bias.
\newblock {\em Epidemiology\/}~{\em 14}, 300--306.

\bibitem[\protect\citeauthoryear{Haavelmo}{Haavelmo}{1943}]{Haavelmo1943}
Haavelmo, T. (1943).
\newblock The statistical implications of a system of simultaneous equations.
\newblock {\em Econometrica\/}~{\em 11}, 1--12.

\bibitem[\protect\citeauthoryear{H\"{u}nermund and Bareinboim}{H\"{u}nermund
  and Bareinboim}{2023}]{Huenermund2021}
H\"{u}nermund, P. and E.~Bareinboim (2023).
\newblock Causal inference and data fusion in econometrics.
\newblock {\em Econom J\/}.
\newblock forthcoming.

\bibitem[\protect\citeauthoryear{Imai, Keele, and Yamamoto}{Imai
  et~al.}{2010}]{Imai2010}
Imai, K., L.~Keele, and T.~Yamamoto (2010).
\newblock Identification, inference and sensitivity analysis for causal
  mediation effects.
\newblock {\em Stat Sci\/}~{\em 25}, 51--71.

\bibitem[\protect\citeauthoryear{Imbens}{Imbens}{2004}]{Imbens2004}
Imbens, G.~W. (2004).
\newblock Nonparametric estimation of average treatment effects under
  exogeneity: A review.
\newblock {\em Rev Econ Stat\/}~{\em 86}, 4--29.

\bibitem[\protect\citeauthoryear{Jones, Molitor, and Reif}{Jones
  et~al.}{2019}]{jones2019}
Jones, D., D.~Molitor, and J.~Reif (2019).
\newblock What do workplace wellness programs do? evidence from the illinois
  workplace wellness study.
\newblock {\em Quart J Econ\/}~{\em 134\/}(4), 1747--1791.

\bibitem[\protect\citeauthoryear{Jung, Tian, and Bareinboim}{Jung
  et~al.}{2021}]{Jung2021}
Jung, Y., J.~Tian, and E.~Bareinboim (2021).
\newblock Estimating identifiable causal effects through double machine
  learning.
\newblock {\em Proc AAAI Conf Artif Intell\/}~(35).

\bibitem[\protect\citeauthoryear{Knaus}{Knaus}{2021}]{Knaus2021}
Knaus, M.~C. (2021).
\newblock A double machine learning approach to estimate efffects of musical
  practice on student's skills.
\newblock {\em J R Stat Soc Ser A Stat Soc\/}~{\em 184\/}(1), 282--300.

\bibitem[\protect\citeauthoryear{Koopmans}{Koopmans}{1947}]{Koopmans1947}
Koopmans, T.~C. (1947).
\newblock Measurement without theory.
\newblock {\em Rev Econ Stat\/}~{\em 29\/}(3), 161--172.

\bibitem[\protect\citeauthoryear{Koopmans}{Koopmans}{1950}]{Koopmans1950}
Koopmans, T.~C. (1950).
\newblock {\em Cowles Foundation Monograph 10: Statistical Inference in Dynamic
  Economic Models}.
\newblock John Wiley \& Sons.

\bibitem[\protect\citeauthoryear{Pearl}{Pearl}{1988}]{Pearl1988}
Pearl, J. (1988).
\newblock {\em Probabilistic Reasoning in Intelligent Systems}.
\newblock San Mateo, CA, USA: Morgan Kaufmann.

\bibitem[\protect\citeauthoryear{Pearl}{Pearl}{1995}]{Pearl1995}
Pearl, J. (1995).
\newblock Causal diagrams for empirical research.
\newblock {\em Biometrika\/}~{\em 82\/}(4), 669--709.

\bibitem[\protect\citeauthoryear{Pearl}{Pearl}{2000}]{Pearl2000}
Pearl, J. (2000).
\newblock {\em Causality: Models, Reasoning, and Inference\/} (1st ed.).
\newblock New York, NY, USA: Cambridge University Press.

\bibitem[\protect\citeauthoryear{Pearl}{Pearl}{2009}]{Pearl2009}
Pearl, J. (2009).
\newblock {\em Causality: Models, Reasoning, and Inference\/} (2nd ed.).
\newblock New York, NY, USA: Cambridge University Press.

\bibitem[\protect\citeauthoryear{Pearl and Mackenzie}{Pearl and
  Mackenzie}{2018}]{Pearl2018}
Pearl, J. and D.~Mackenzie (2018).
\newblock {\em The Book of Why: The New Science of Cause and Effect}.
\newblock New York: Basic Books.

\bibitem[\protect\citeauthoryear{Peters, Janzing, and Sch\"{o}lkopf}{Peters
  et~al.}{2017}]{Peters2017}
Peters, J., D.~Janzing, and B.~Sch\"{o}lkopf (2017).
\newblock {\em Elements of Causal Inference}.
\newblock Cambridge, MA, USA: MIT Press.

\bibitem[\protect\citeauthoryear{Robinson}{Robinson}{1988}]{Robinson1988}
Robinson, P.~M. (1988).
\newblock Root-n-consistent semiparametric regression.
\newblock {\em Econometrica\/}~{\em 56\/}(4), 931--954.

\bibitem[\protect\citeauthoryear{Spirtes, Glymour, Scheines, and
  Heckerman}{Spirtes et~al.}{2000}]{Spirtes2000}
Spirtes, P., C.~N. Glymour, R.~Scheines, and D.~Heckerman (2000).
\newblock {\em Causation, Prediction, and Search}.
\newblock Cambridge, MA, USA: MIT Press.

\bibitem[\protect\citeauthoryear{Strotz and Wold}{Strotz and
  Wold}{1960}]{Strotz1960}
Strotz, R.~H. and H.~O.~A. Wold (1960).
\newblock Recursive vs.\ nonrecursive systems: An attempt at synthesis (part i
  of a triptych on causal chain systems).
\newblock {\em Econometrica\/}~{\em 28}, 417--427.

\bibitem[\protect\citeauthoryear{Tibshirani}{Tibshirani}{1996}]{Tibshirani1996}
Tibshirani, R. (1996).
\newblock Regression shrinkage and selection via lasso.
\newblock {\em J R Stat Soc Series B Stat Methodol\/}~{\em 58\/}(1), 267--288.

\bibitem[\protect\citeauthoryear{Vanneste and Gulati}{Vanneste and
  Gulati}{2021}]{Vanneste2021}
Vanneste, B.~S. and R.~Gulati (2021).
\newblock Generalized trust, external sourcing, and firm performance in
  economic downturns.
\newblock {\em Organ Sci\/}~{\em 33\/}(4), 1251--1699.

\bibitem[\protect\citeauthoryear{Woodward}{Woodward}{2003}]{Woodward2003}
Woodward, J. (2003).
\newblock {\em Making Things Happen}.
\newblock Oxf Stud Philos Sci. Oxford, UK: Oxford University Press.

\bibitem[\protect\citeauthoryear{W\"{u}thrich and Zhu}{W\"{u}thrich and
  Zhu}{2021}]{Wuethrich2021}
W\"{u}thrich, K. and Y.~Zhu (2021).
\newblock Omitted variable bias of lasso-based inference methods: A finite
  sample analysis.
\newblock {\em Rev Econ Stat\/}.
\newblock forthcoming.

\end{thebibliography}


\end{document}